\documentclass[aip,rsi,superscriptaddress,citeautoscript]{revtex4-1}

\usepackage{graphicx}
 \usepackage{epstopdf}
\usepackage{dcolumn}%
\usepackage{bm}%
\usepackage{amsmath}
\usepackage[english]{babel}

\draft 

\begin{document}

\preprint{submitted to Review of Scientific Instruments}

\title{Background-free quasi-steady-state photoinduced absorption spectroscopy by dual optical modulation} 



\author{Houssem~Kallel}
\affiliation{D\'{e}partement de physique et Regroupement qu\'{e}b\'{e}cois sur les mat\'{e}riaux de pointe, Universit\'{e} de Montr\'{e}al,  C.P.\ 6128, Succursale centre-ville, Montr\'{e}al (Qu\'{e}bec), H3C~3J7, Canada}

\author{Gianluca~Latini}
\affiliation{Department of Materials and Centre for Plastic Electronics, Imperial College London, South Kensington Campus, London SW7~2AZ, United Kingdom}

\author{Francis~Paquin}
\affiliation{D\'{e}partement de physique et Regroupement qu\'{e}b\'{e}cois sur les mat\'{e}riaux de pointe, Universit\'{e} de Montr\'{e}al,  C.P.\ 6128, Succursale centre-ville, Montr\'{e}al (Qu\'{e}bec), H3C~3J7, Canada}

\author{Robert~Rinfret}
\affiliation{D\'{e}partement de physique et Regroupement qu\'{e}b\'{e}cois sur les mat\'{e}riaux de pointe, Universit\'{e} de Montr\'{e}al,  C.P.\ 6128, Succursale centre-ville, Montr\'{e}al (Qu\'{e}bec), H3C~3J7, Canada}

\author{Natalie~Stingelin}
\affiliation{Department of Materials and Centre for Plastic Electronics, Imperial College London, South Kensington Campus, London SW7~2AZ, United Kingdom}

\author{Carlos~Silva}
\affiliation{D\'{e}partement de physique et Regroupement qu\'{e}b\'{e}cois sur les mat\'{e}riaux de pointe, Universit\'{e} de Montr\'{e}al,  C.P.\ 6128, Succursale centre-ville, Montr\'{e}al (Qu\'{e}bec), H3C~3J7, Canada}
\email[corresponding author. E-mail address:  ]{carlos.silva@umontreal.ca}
\homepage[Website: ]{http://www.phys.umontreal.ca/~silva/}


\date{\today}

\begin{abstract}
We present a simple and inexpensive means to measure background-free photoinduced absorption spectra by implementing a dual optical modulation scheme. Our objective is to measure quasi-steady-state absorption spectra of long-lived photoexcitations, such as polarons and triplet excitons in organic semiconductors, when the photoluminescence of the sample is strong compared to the photoinduced absorption signal. In our instrument, we modulate a continuous-wave laser at a frequency $f_{pu}$ with a mechanical chopper before exciting the material. We measure the fractional change of transmission of the sample with a tuneable monochromatic probe source modulated at a frequency $f_{pr}$, and with a digital lockin amplifier at a reference frequency $f_{ref} = f_{pu} + f_{pr}$. We generate the reference sinusoidal wave for the lockin amplifier using a simple home-built electronic circuit that generates their sum-frequency signal. This scheme allows measurement of the fractional change of transmittance of the sample induced by the pump laser at a frequency $f_{pu} + f_{pr}$ without parasitic signal at a frequency $f_{pu}$, such as photoluminescence of the sample or pump-laser scatter. We demonstrate the power of this approach by measuring cleanly the photoinduced absorption spectrum of a highly luminescent semiconductor polymer, poly(9,9-dioctylfluorene-co-benzothiadiazole), at 10\,K. We compare this measurement favourably  with that obtained by simple modulation of the pump laser after subtraction of the dominant photoluminescence background.
\end{abstract}


\maketitle 

\section{Introduction}
\label{Intro}
Modulation spectroscopy is a powerful means of characterizing photoexcitations in semiconductors, and has been exploited in numerous variations over four decades~\cite{Cardona:1969bh}. Following the development of digital lockin amplifiers in the 1980s, modulation spectroscopies have been used extensively to measure minuscule steady-state signals following continuous-wave photoexcitation. Photoinduced absorption~\cite{PhysRevB.42.11670,PhysRevLett.72.156} and photoreflectance~\cite{Glembocki:1992fk} spectroscopies are two important examples of such techniques. In these, a sample of interest is excited resonantly with a continous-wave (CW) optical source, usually a laser, that is modulated either with a mechanical chopper or with an acousto-optic modulator, depending of the desired frequency range. A tuneable monochromatic probe light source is then used to measure the photoinduced changes in absorbance or reflectance induced by the photomodulated pump source with the aid of a lockin amplifier. Such techniques measure the quasi-steady-state photoinduced population change at the pump modulation frequency and the probe photon energy, which is then scanned to construct the photomodulation spectrum. These techniques are very powerful, because they permit the examination of the electronic structure of materials, as well as decay kinetics of photoinduced species (e.g.\ carriers in inorganic semiconductor nanorstructures~\cite{SHANABROOK:1987fk}, or polarons and triplet excitons in organic semiconductor heterostructures~\cite{Wohlgenannt:1999ul}), by analysing the frequency dependence of the excitation laser. However, because this technique measures quasi-steady-state population, which is usually very low relative to absorption cross sections (for example, $\sim 10^{-16}$\,cm$^2$ for triplet exciton absorption in semiconductor polymers~\cite{Lee:2007qf}), photoinduced absorption signals are typically very small, with typical fractional changes in transmittance of the sample between $10^{-3}$ and $10^{-5}$, requiring phase-sensitive detection techniques. In this paper we focus on photoinduced absorption spectroscopy, but our apparatus is equally suitable for photoreflectance spectroscopy.

A complication in photoinduced absorption measurements arises when the sample of interest displays efficient photoluminescence (PL) over a similar spectral range as the desired signal, which may mask it altogether. It is in principle possible to subtract the PL contribution to yield the photoinduced absorption signal only, but such procedure is time-consuming because sufficient signal-to-noise ratio is required to extract a small signal discussed above from the dominant PL signal, which may be orders of magnitude more intense. For example, the PL quantum efficiency of F8BT, the semiconductor polymer that we will examine in section~\ref{sec:F8BT}, is 60--70\% at room temperature~\cite{*[{See Table 1 in }] [{ for a summary of PL quantum efficiency values published for F8BT.}] Sardone:2007fk}. Considering the geometry of the experimental setup, the signal from PL background to photoinduced absorption can be $10^{-1}$ -- $10^{-2}$, which can be four orders of magnitude higher than the signal of interest. In order to resolve this issue, we implement a double modulation scheme in which the probe source is modulated at a different frequency to the pump, and we analyze the signal at the sum-frequency of the pump and probe beams. The signal at this sum frequency is free of PL background and other parasitic components such as scatter from the pump source. We demonstrate that it is possible to measure cleanly photoinduced absorption spectra of highly luminescent materials. In addition, by simultaneously analyzing the signal at the probe modulation frequency only, we can measure the ground-state absorption spectrum independently. This apparatus permits the measurement of a fractional change in transmittance of $\sim 10^{-6}$ in our implementation, currently limited by the temperature stability of our grey-body probe source, and by the shot noise of the detector. Our instrument is also well suited for photoreflectance spectroscopy.

\section{Instrument}
\subsection{Quasi-steady-state photoinduced absorption spectroscopy}
\label{subsec:PIA}

In the photoinduced absorption experiment, a CW pump laser, modulated at a frequency $f_{pu}$, excites the sample and the fractional change in the transmission of a probe beam through the sample, $\Delta T / T$, is measured. By the Beer-Lambert law, this signal is related to the density of the photoinduced (absorbing) species, $n$, by~\cite{Silva:2002dq}
\begin{equation}
-n \sigma z =  \ln \left (1 + \frac{\Delta T}{T} \right ) \approx  \frac{\Delta T}{T}
\label{BeersLaw}
\end{equation}
if $\Delta T / T \ll 1$. In equation~\ref{BeersLaw}, $\sigma$ is the absorption cross section and $z$ is the sample thickness. In a general scenario, the long-lived photoexcitation dynamics of the photoinduced species can decay with a unimolecular rate constant $\gamma$ and a bimolecular rate constant $\beta$, such that
\begin{equation}
\label{eq:piaDynamics}
\frac{dn}{dt} = \kappa \rho G(t) - \gamma n - \beta n^{2},
\end{equation}
where $\kappa$ is the photogeneration efficiency (number of species generated per absorbed photon), $\rho$ is the rate of photon absorption per unit volume, and $G$ is a function describing the pump modulation.  Equation~\ref{eq:piaDynamics} does not yield a unique lifetime that is independent of laser intensity with a constant value during the entire relaxation process.  However, let us consider $n_s(\omega)$, the component of $n(t)$, the actual solution, at the fundamental  frequency $\omega = 2 \pi f_{pu}$ and under quasi-steady-state conditions. Let us assume that  $G$ adopts a symmetric square-wave functional form. Under these assumptions, the steady-state population density is given by~\cite{DELLEPIANE:1993mz}
\begin{eqnarray}
\label{eq:piaSoln}
n_{s}(\omega) &=& \frac{\kappa \rho \tanh \frac{\gamma \pi}{2 \omega} \tanh \frac{P \pi}{\omega}}{\frac{\gamma}{2} \tanh \frac{P \pi}{\omega} + P \tanh \frac{\gamma \pi}{2 \omega}}; \\
P &=& \sqrt{\beta \kappa \rho + \frac{\gamma^{2}}{4}}.
\label{eq:n_s}
\end{eqnarray}

 \begin{figure}
 \includegraphics[scale=1]{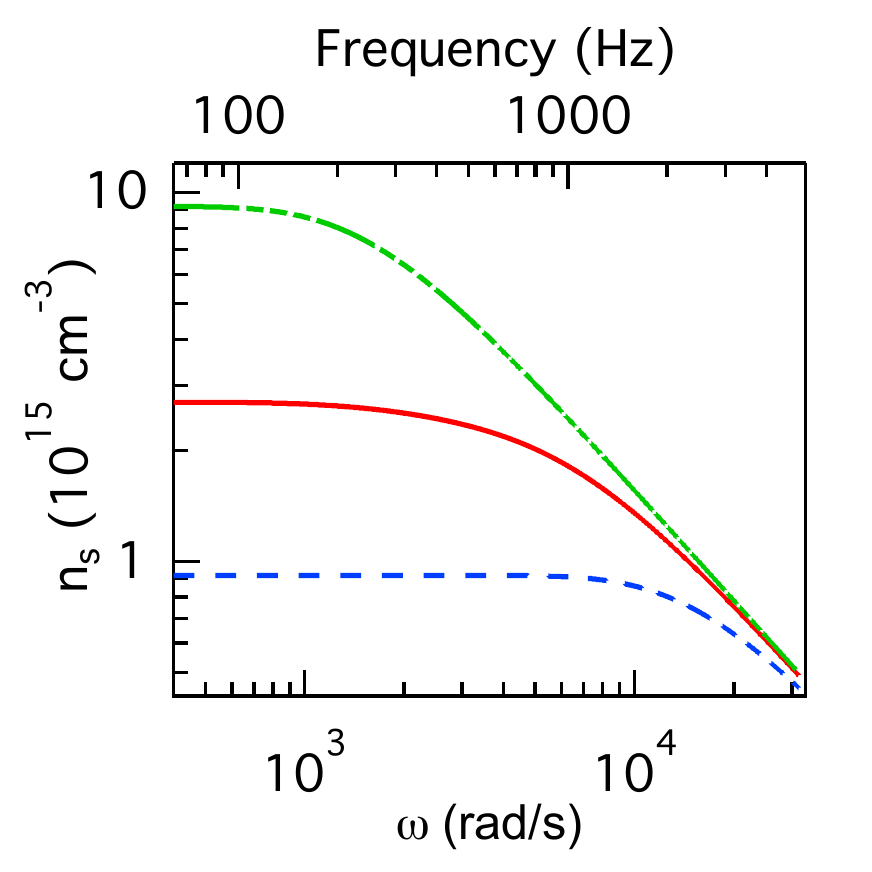}
 \caption{Steady-state population density versus pump-laser modulation frequency according to equation~\ref{eq:piaSoln}, with $\kappa \rho = 1 \times 10^{19}$\,cm$^{-3}$s$^{-1}$ for all traces. Shown are the predicted frequency dependence when $\gamma = 1 \times 10^{3}$\,s$^{-1}$, and $\beta = 1 \times 10^{-12}$\,cm$^{3}$s$^{-1}$ (red, continuous line); $\gamma = 1 \times 10^{3}$\,s$^{-1}$, and $\beta = 1 \times 10^{-14}$\,cm$^{3}$s$^{-1}$ (green, dash-dotted  line); and $\gamma = 1 \times 10^{4}$\,s$^{-1}$, and $\beta = 1 \times 10^{-12}$\,cm$^{3}$s$^{-1}$ (blue, dashed line).}
 \label{fig1}
 \end{figure}
Fig.~\ref{fig1} displays the dependence of $n$ with frequency as predicted by equation~\ref{eq:piaSoln}, using values for $\kappa$, $\gamma$, and $\beta$ that are typical for polarons and triplet excitons in neat semiconductor polymers (see, for example, ref.~\onlinecite{Silva:2002dq}). In general, $n$ is weakly dependent on frequency at low modulation frequency, but falls off as $\omega^{-1}$ at high frequency. When the modulation period is long compared to the average decay time constant defined by $\gamma$ and $\beta$, $n$ is in quasi steady state. As the modulation frequency is increased and the modulation period is comparable to the average decay time constant, the population density becomes dependent on modulation frequency. Thus, the threshold frequency for which $n$ depends on frequency, $f_{0}$, is indicative of the average decay time constant, and it general, it increases with $\beta$ and $\gamma$.~\footnote{from equation~\ref{eq:n_s}, the threshold value of $\omega$ in which $n_s$ changes from the limit of $\omega \rightarrow 0$ to $\omega \rightarrow \infty$ occurs at $\omega_0 = \left(\gamma \pi / 4 \right) \left(1 + \sqrt{1 + (4 \beta \kappa \rho / \gamma^2)}\right)$. Thus, $\omega_0$ increases with both $\beta$ and $\gamma$.} Fig.~1 compares the effect of varying $\gamma$ and $\beta$ on $f_{0}$. Upon increasing $\gamma$ by an order of magnitude from $1 \times 10^{3}$\,s$^{-1}$ (red, continuous line versus blue, dashed line), the quasi-steady-state population density decreases due to the higher rate of population depletion, but $f_{0}$ increases from $\sim 650$\,Hz to $\sim 1.8$\,kHz. Similarly, decreasing $\beta$ by a factor of 100 from $1 \times 10^{-12}$\,cm$^{3}$s$^{-1}$ to $1 \times 10^{-14}$\,cm$^{3}$s$^{-1}$ (green, dash-dotted line versus blue, dashed line) also results in a drop in $n$, and $f_0$ decreases to $\sim 200$\,Hz. In order to operate close to steady state in a wide range of situations, we therefore choose a modulation frequency of $f_{pu} = 170$\,Hz for this experimental design, which generally places us in a regime of weak frequency dependence of $n$ according to Fig.~\ref{fig1}.

\subsection{Double optical modulation approach}
\label{subsec:DoubleMod}

In the double modulation approach, we modulate the laser excitation beam at a frequency $f_{pu}$ as in the standard photoinduced absorption geometry described above, but we also modulate the probe beam at a frequency $f_{pr} \neq f_{pu}$.  In such configuration, the output signal contains components at the individual frequencies  $f_{pu}$ and $f_{pr}$, but also at the sum frequency $f_{pu} + f_{pr}$ and the difference frequency $f_{pu} - f_{pr}$. We show below that the ratio $\left (\Delta I_{out}^{f_{pu}+f_{pr}} / \Delta I_{out}^{f_{pr}} \right )$ between the amplitude of the component at the sum frequency and the one at  the probe frequency is identical to the signal from the conventional single modulation configuration $\left (\Delta I_{out}^{f_{pu}} / I_{out}^{dc} \right )$. This indicates that the spectral features measured with the two methods are identical. In order to show this we take into account the fact that the transmission is a function of the intensity of the pump excitation $I_{pu}$ only, so that the intensity of the transmitted light is given by
\begin{equation}
I_{out}(I_{pu},I_{pr}) = I_{pr} \cdot T(I_{pu}),
\label{eq:Iout0}
\end{equation}
where $I_{pr}$ is the intensity of the incident probe beam. Both $I_{pu}$ and $I_{pr}$ are modulated functions, and considering only the first harmonic (sine term) in a Fourier series expansion of the modulation function,
\begin{subequations}
\begin{align}
I_{pu} &= \frac{I^0_{pu}}{2} \left[ 1 + \sin (2 \pi f_{pu} t) \right] \equiv  \frac{I^0_{pu}}{2} + \Delta I_{pu}; \\
I_{pr} &= \frac{I^0_{pr}}{2} \left[ 1 + \sin (2 \pi f_{pr} t) \right] \equiv  \frac{I^0_{pr}}{2} + \Delta I_{pr} .
\end{align}
\label{eq:I}
\end{subequations}
The photoinduced variation of the transmitted light  $\Delta I_{out}$ can be calculated by Taylor-expanding the differential of $I_{out}$  up to second order around the average values  $I^0_{pu} / 2$ and  $I^0_{pr} / 2$.  This gives
\begin{align}
\Delta I_{out} = & \frac{I^0_{pr}}{2} \left[ \frac{\mathrm{d}T}{\mathrm{d}I_{pu}} \right]_{I^0_{pu}/2} \Delta I_{pu} + T\left(\frac{I^0_{pu}}{2}\right) \Delta I_{pr} \nonumber \\
 &+ \frac{1}{2}  \frac{I^0_{pr}}{2} \left[ \frac{\mathrm{d}^2T}{\mathrm{d}I_{pu}^2} \right]_{I^0_{pu}/2}\Delta I_{pu}^2
 + \left[ \frac{\mathrm{d}T}{\mathrm{d}I_{pu}} \right]_{I^0_{pu}/2}\Delta I_{pu}\Delta I_{pr} + \cdots
 \label{eq:Iout1}
\end{align}
where we have considered that $T$ is a function of $I_{pu}$ only. Thus, from equations~\eqref{eq:I}, and invoking the trigonometric identity $2 \sin \alpha \sin \beta = \cos(\alpha - \beta) - \cos(\alpha + \beta)$, we obtain
\begin{align}
\Delta I_{out} = & \frac{I^0_{pu}I^0_{pr}}{4} \left[ \frac{\mathrm{d}T}{\mathrm{d}I_{pu}} \right]_{I^0_{pu}/2} \sin(2 \pi f_{pu} t) \nonumber \\
& +  \frac{I^0_{pr}}{2} \cdot T\left( \frac{I^0_{pu}}{2} \right) \sin(2 \pi f_{pr} t) \nonumber \\
& + \frac{1}{2} \frac{\left(I^0_{pu}\right)^2 I^0_{pr}}{8}  \left[ \frac{\mathrm{d}^2T}{\mathrm{d}I_{pu}^2} \right]_{I^0_{pu}/2} \sin^2(2 \pi f_{pu} t) \nonumber \\
& + \frac{I^0_{pu} I^0_{pr}}{8} \left \{ \cos \left[2 \pi (f_{pu} - f_{pr}) t \right] - \cos \left[2 \pi (f_{pu} + f_{pr}) t \right] \right \} + \cdots
\label{eq:Iout2}
\end{align}
In the double modulation configuration, we measure the pump-induced signal at the sum of the pump and probe modulation frequencies, $\Delta I_{out}^{f_{pu}+f_{pr}}$, normalized by the signal at the probe modulation frequency, $\Delta I_{out}^{f_{pr}}$:
\begin{equation}
\frac{\Delta I_{out}^{f_{pu}+f_{pr}}}{\Delta I_{out}^{f_{pr}}} = \frac{I^0_{pu}}{4} \frac{\left[ \mathrm{d}T / \mathrm{d}I_{pu} \right]_{I^0_{pu}/2}}{T \left(I^0_{pu} / 2 \right)}.
\end{equation}
In a single modulation scheme, equation~\eqref{eq:Iout1} still applies provided that $\Delta I_{pr} = 0$, so when the series is calculated around $I_{pu} = I^0_{pu}/2$ for $I_{pr} = I^0_{pr}$ (in this case $I_{pr}$ is not modulated) then from equations~\eqref{eq:Iout2} and \eqref{eq:Iout0} we obtain
\begin{equation}
\frac{\Delta I_{out}^{f_{pu}}}{I_{out}^{dc}} = \frac{I^0_{pu}I^0_{pr}}{4} \frac{\left[ \mathrm{d}T / \mathrm{d}I_{pu} \right]_{I^0_{pu}/2}}{I^0_{pr} T \left(I^0_{pu} / 2 \right)} 
= \frac{\Delta I_{out}^{f_{pu}+f_{pr}}}{\Delta I_{out}^{f_{pr}}}.
\end{equation}
The output signal in single and double modulation is therefore identical.

We seek to detect the component of the photoinduced absorption signal at the sum of the two modulation frequencies $f_{pu} +f_{pr}$. Let us now consider an optimal probe frequency necessary to meet the most stringent requirements of background-free detection. The probe beam is aligned through a mechanical chopper prior to being transmitted through the sample and being detected by a photoreceiver. We can idealize the time-dependent signal of the modulated probe beam at the detector as a square-wave function of period $2 \tau$ and amplitude $V_0$. Over the period $[0,2\tau]$, this can be written as
\begin{equation}
g_{pr}(t) = V_0 \left[ H\left(\frac{t}{\tau}\right) - H\left(\frac{t}{\tau} - 1\right) \right],
\label{eq:squarewav}
\end{equation}
where $H$ is the heaviside step function, which, if expressed as a piecewise constant function, is defined as
\begin{equation}
H(x) = 
\begin{cases} 
   0 & \text{if } x < 0 \\
   \frac{1}{2} & \text{if } x = 0 \\
   1 & \text{if } x > 0
  \end{cases}
  .
\end{equation}
The Fourier-series expansion of $g_{pr}(t)$ contains only odd terms:
\begin{equation}
g_{pr}(t) = V_0 \left[ \frac{1}{2} +\frac{2}{\pi} \sum_{m = 1,3,5,\cdots}^{\infty} \frac{1}{m} \sin \left(\frac{m \pi t}{\tau} \right) \right].
\end{equation}
Low odd harmonics $m$ are therefore expected to contain significant amplitude compared to the signal of interest, and their `floor' noise could be substantial. In order to achieve truly background-free signal and high sensitivity, we must choose our modulation frequencies such that $f_{pu} +f_{pr}$ is sufficiently different from $f_{pr}$ or one of its first few odd harmonics. We choose $f_{pr} = 139$\,Hz, placing the desired photoinduced absorbance signal at $f_{pu} +f_{pr} = 309$\,Hz, which is sufficiently far from $3f_{pr} = 417$\,Hz and should therefore not be affected by floor noise of the intense third harmonic of the probe modulation frequency. 

 \begin{figure}
 \includegraphics[scale=1]{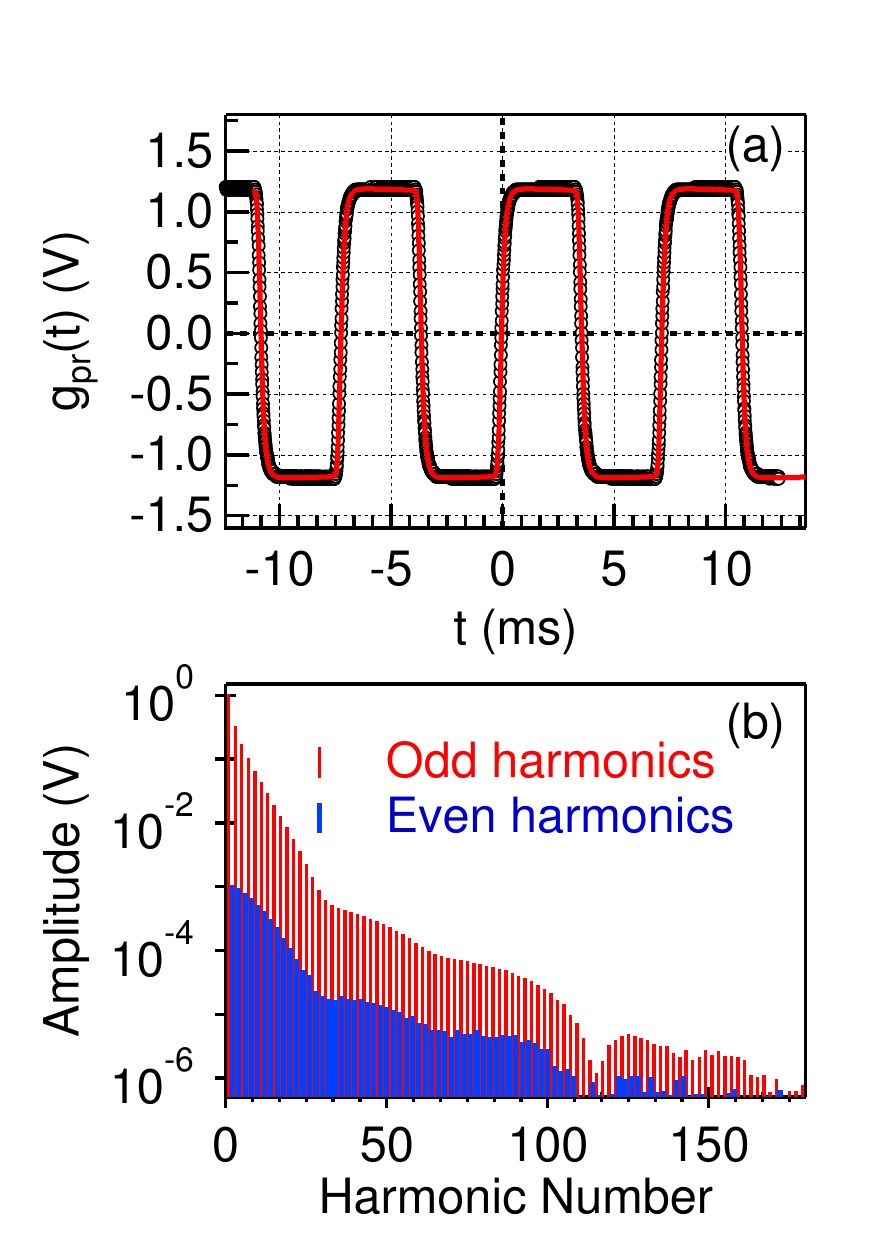}
 \caption{(a) Measured response of the photodiode upon modulation of the probe lamp at a frequency of 139\,Hz (open circles), measured at 700\,nm. (b) Amplitude of the signal from the lockin amplifier as a function of harmonic number. The red, continuous line in part (a) is the reconstructed response function from the amplitudes in part (b) as the coefficients of a Fourier series.}
 \label{fig2}
 \end{figure}
 
 Fig.~\ref{fig2}(a) displays the measured form of $g_{pr}(t)$ (open circles) obtained by modulating the output of a tungsten-halogen lamp at a frequency of 139\,Hz using a digital oscilloscope. This signal deviates from an ideal square-wave function expressed by equation~\ref{eq:squarewav}. We have measured this response at various wavelengths and find that it does not depend on probe wavelength. In order to characterize the experimental form of $g_{pr}(t)$, we have measured the amplitude of the signal using a digital lockin amplifier (SRS SR830), for which the reference input was the TTL output of the chopper controller at 139\,Hz, as a function of harmonic of the reference signal.~\footnote{Following this procedure, the component of the amplitude of the signal at each harmonic can be measured independently, up to harmonic number 733 at a fundamental frequency of 139\,Hz with the SR830 lockin amplifier. The SR830 lockin amplifier sets the reference signal from its internal oscillator to a user-defined harmonic up to a maximum frequency of 102.4\,kHz.} The results are displayed in Fig~\ref{fig2}(b) up to the 180th harmonic. We observe that even harmonics (blue bars) have amplitude of $\sim 10^{-3}$ times that of odd harmonics (red bars), but may be of comparable amplitude to the desired signal level at frequency $f_{pu} +f_{pr}$. At the chosen probe modulation frequency, $2f_{pr} = 278$\,Hz, which is closer to the sum frequency at 309\,Hz. We address this in more detail in the following section. We have reconstructed $g_{pr}(t)$ using the harmonic amplitudes as Fourier components, and we display it as a red continuous line in part (a). 

\subsection{Experimental apparatus}
 
 \begin{figure}
 \includegraphics[width=\textwidth]{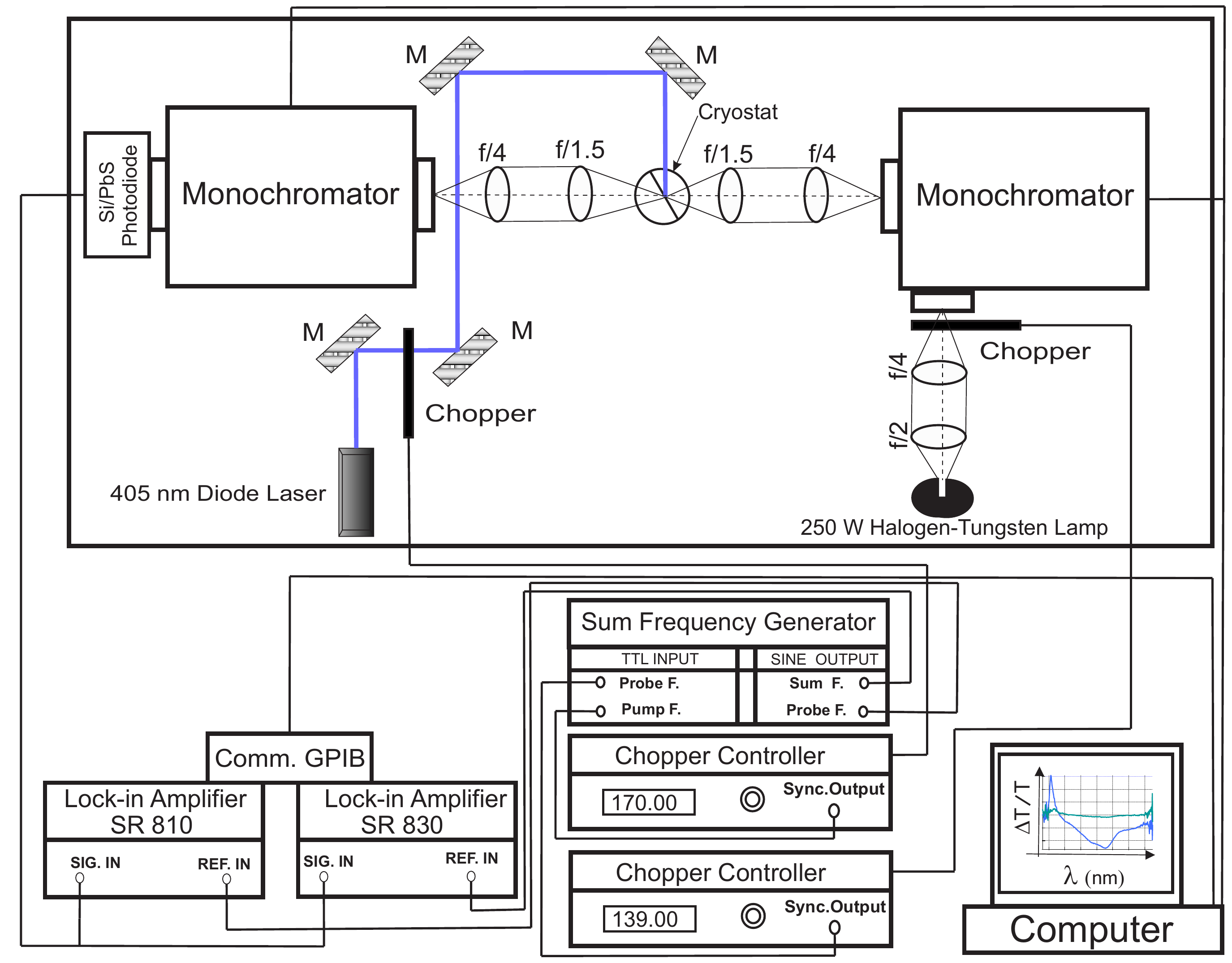}
 \caption{Schematic diagram of the experimental instrument.}
 \label{fig3}
 \end{figure}
 We now discuss the experimental apparatus, shown schematically in Fig.~\ref{fig3}. The sample is contained in a cryostat (Cryo Industries, closed cycle model 350 optical refrigerator cryostat). To pump the sample, a semiconductor laser (Newport, $\lambda = 405$\,nm, 50\,mW) was modulated by an optical chopper (Terahertz Technologies model C-995) at frequency $f_{pu} = 170$\,Hz and directed to the sample. The probe source was a 250-W tungsten-halogen lamp (Oriel), coupled to a 300-mm focal length monochromator (PI-Acton SP-2300i). The output of this monochromator was modulated at frequency $f_{pr} = 139$\,Hz by placing an identical chopper to that for the pump laser at the entrance slit of the monochromator. The monochromatic output was then imaged through the sample (1/e$^2$ beam waist of $178 \pm 30$\,$\mu$m at the sample), where it is overlapped with the pump laser (1/e$^2$ beam waist of $989 \pm 17$\,$\mu$m), onto the entrance slit of a spectrometer with focal length 300\,mm (PI-Acton SP-2300i), which was matched with the first monochromator. A photoreceiver consisting of a peltier-cooled sandwich diode structure of Si and PbS (Electro-Optical Systems S/PBS-025/020-TE2-H) detected the probe light.~\footnote{The bandwidth of the S/PBS-025/020-TE2-H photoreceiver in low gain mode is 2\,kHz for the silicon diode and that of the lead sulphide detector is 500\,Hz. Therefore, in order to take advantage of the optimal responsivity of the device, modulation frequencies $f_{pu}, f_{pr} < 200$\,Hz are desirable.} The output of this receiver was then split and sent to the input of two digital lockin amplifiers (Stanford Research Systems SR830 and SR810). We constructed an electronic circuit to generate the reference signals for the two lockin amplifiers. The inputs of this circuit are the TTL sync output of the two chopper controller units, at frequencies 170 and 139\,Hz. Its outputs are two sine waveforms, one at 309\,Hz, the sum frequency, and one at 139\,Hz. The former was used as a reference on the dual-phase lockin amplifier (SR830), which serves to measure the fractional change in transmittance induced by the pump laser ($\Delta T$), while the second served that purpose on the single-phase amplifier (SR810), measuring the probe transmittance ($T$) through the sample. The lockin amplifiers, the spectrometers and the cryostat are interfaced to a computer, so that the spectral response of $\Delta T/T$ can be measured as a function of temperature over a spectral range spanning 500 and 2800\,nm, and over a temperature range ranging from 10\,K to room temperature.

\begin{figure}
 \includegraphics[width=\textwidth]{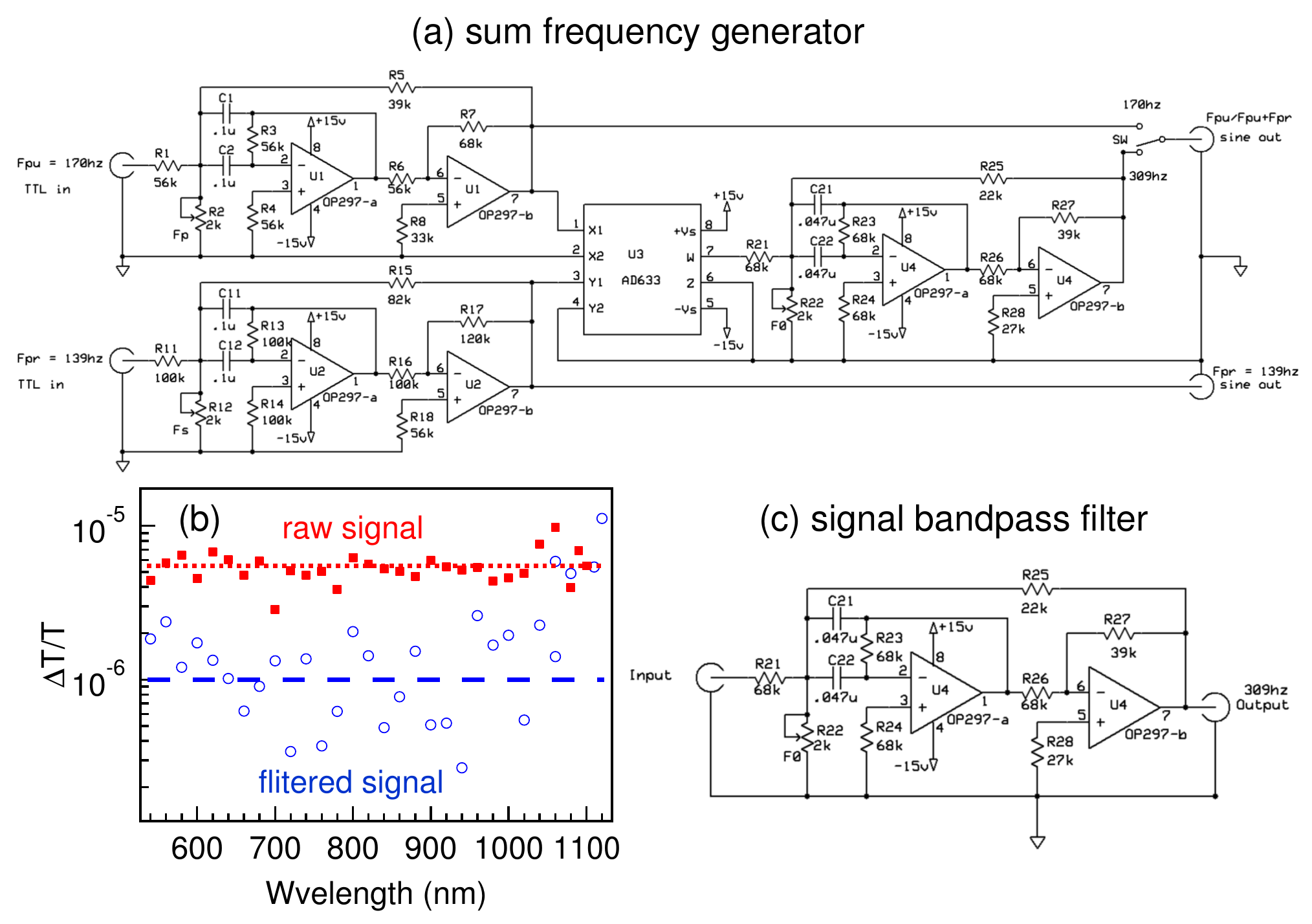}
 \caption{(a) Schematic diagram of the sum-frequency generation electronic circuit. Also shown is the $\Delta T / T$ baseline signal (b) when the Si photoreceiver output is filtered (blue open circles) and unfiltered (solid red squares) with the circuit shown in part (c). The spectra were collected with a time constant of 300\,ms, and two spectral scans. The blue, dashed line in part (b) is at $\Delta T / T = 1 \times 10^{-6}$, and the red, dotted line is at $\Delta T / T = 5 \times 10^{-6}$.}
 \label{fig4}
 \end{figure}
 A schematic diagram of the sum-frequency generation circuit is shown in Fig.~\ref{fig4}(a). Similar circuits have been published by Ghosh and Arora~\cite{*[{The circuit reported in Fig.~2 of }] [{ has a few connections omitted from the schematic diagram and therefore does not function as drawn. Our circuit is similar to the one reported in that paper, but it implements a newer analog frequency mixing chip and is simpler.}] Ghosh:1998zr} and by Ho~et~al.~\cite{*[{The circuit shown in Fig.~2 of }] [{ implies that TTL signals can be summed by an AD633 analog multiplier (also used in our circuit), which is incorrect. This multiplier requires sinusoidal inputs, and does not function with TTL pulsetrains as drawn in the paper by Ho~et~al.}] Ho:2001gf} Both input TTL waveforms are converted to sinusoidal signals by bandpass filters ($Q = 40$) at 170 and 139\,Hz for the pump and probe channels, respectively. Both channels are then input to an AD633 four-quadrant analog multiplier. The sinusoidal signal at the sum frequency is then passed through a $Q = 40$ bandpass filter at 309\,Hz, and is used as a reference signal for one lockin amplifier. We also isolate the component at 139\,Hz and use it as a reference for the other lockin amplifier. It is possible to bypass the sum-frequency generation process in the circuit with a switch in order to produce a sinusoidal output at 170\,Hz and thereby switch rapidly between single and double modulation experiments if desired.
 
 We find that the instrument spectral response at the reference frequency of 309\,Hz contains a small but non-negligible contribution of the probe lamp, shown as red, filled squares in  Fig.~\ref{fig4}(b). The signal in the near-infrared measured with the PbS photoreceiver shows this artifactual signal more prominently (not shown here), as the probe lamp intensity is higher in that spectral range. This is presumably due to floor-noise contributions from the second harmonic of $f_{pr}$ (at 278\,Hz). In order to remove this undesired background response, we filter the photoreceiver output using a 309-Hz bandpass filter, shown schematically in Fig.~\ref{fig4}(c), before the input of the SR830 lockin amplifier (blue open circles in part (b)). This produces a baseline response free of the lamp response, with baseline at $\Delta T / T = 1 \times 10^{-6}$.
 
 The main advantage of this experimental scheme is that it permits essentially background-free detection of the $\Delta T/T$ spectrum. In addition, in our implementation we detect $\Delta T$ and $T$ \emph{simultaneously}, as opposed to a single modulation approach in which a spectrum of $\Delta T$ and $T$ need to be collected separately by moving the chopper from the pump arm to the probe arm and repeating the measurement, if the highest signal-to-noise ratio is sought. In a single modulation experiment with a high luminescence background, three experiments therefore need to be performed. Two in which the pump laser is modulated and the probe beam is either blocked or unblocked to permit subtraction of the PL background from the desired $\Delta T$, and one in which the probe beam is modulated and the pump laser is blocked, to measure $T$. We consider that the clear simplicity of the double modulation approach is a significant advantage.
 
 We note that it would also be possible to implement a double modulation scheme by using a mechanical chopper (such as the SRS SR540) with two different sets of slots in order to modulate independently two optical beams at different but related frequencies. The controller unit of that chopper is capable of supplying TTL sync outputs at the two modulation frequencies, which would permit implementation of similar signal processing with two lockin amplifiers. We opt for the approach implementing the sum-frequency generation circuit described in this paper because it permits a simpler optical alignment, and because it permits us to choose two unrelated modulation frequencies using the logic discussed in Section~\ref{subsec:DoubleMod}. We also note that Kita~et~al.\ reported a double modulation scheme for photoreflectance measurements in inorganic semiconductor nanostructures with a function generator to supply two trigger signals to independent choppers, and a reference signal at the sum frequency to a lockin amplifier~\cite{Kita:2008fr}. We compared this approach to the sum-frequency generation circuit discussed here, and found that the phase stability was superior with the latter approach, even when a high-stability multichannel function generator was used (Berkeley Nucleonics model 575 digital delay/pulse generator) .
 
 Finally, we point out that because the signal at frequency $f_{pu} + f_{pr}$ is free of PL background, one could implement the experiment without the detection spectrometer in Fig.~\ref{fig3}, which would  simplify the experimental setup. We choose to retain it because the setup as shown here allows us to measure PL spectra by blocking the probe beam and switching the reference frequency output of our home-built circuit to $f_{pu}$.

\section{Experimental demonstration}
\label{sec:F8BT}

\begin{figure}
 \includegraphics[scale=1]{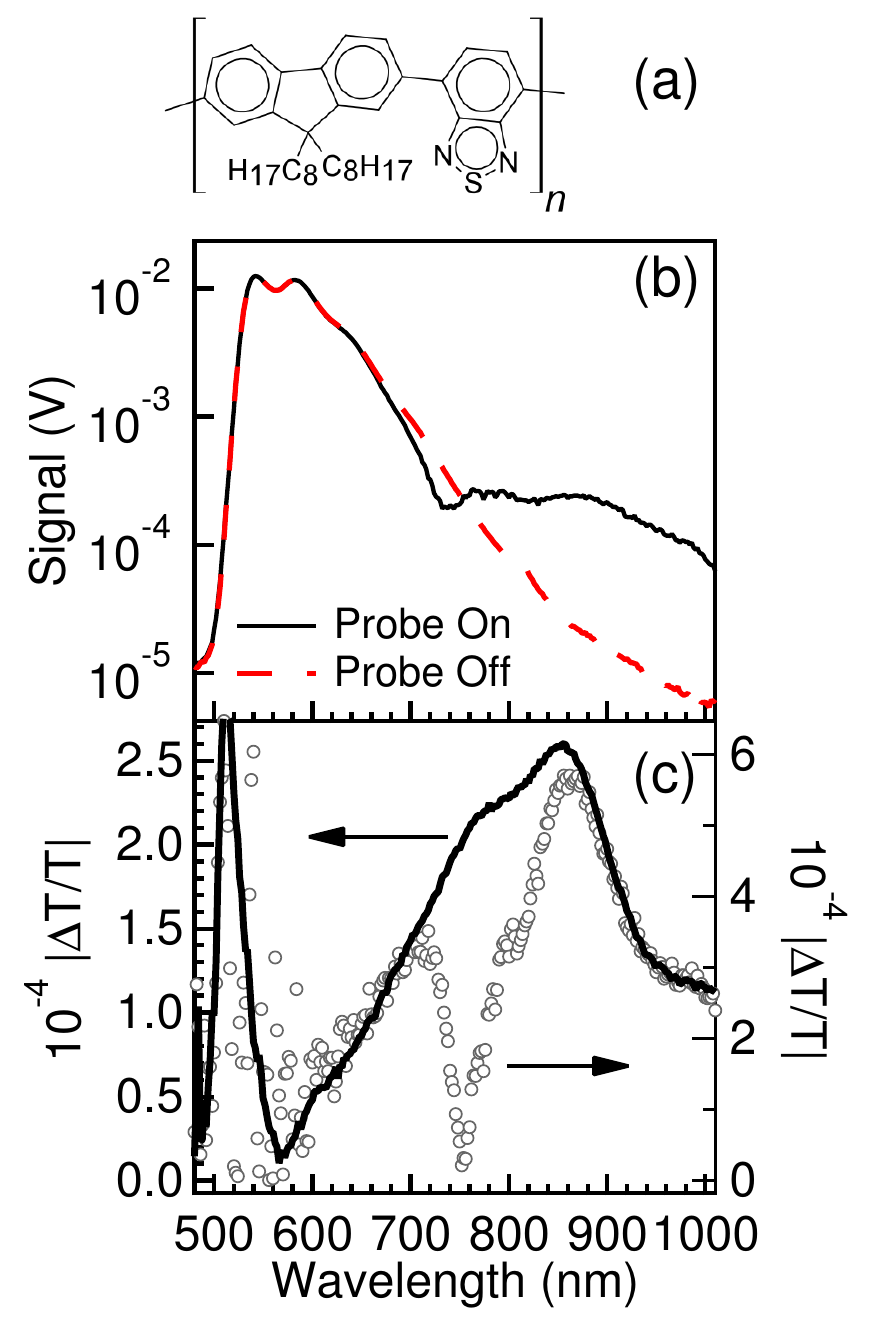}
 \caption{(a) Molecular structure of F8BT. (b) Signal obtained at 10\,K by modulating the pump laser at a frequency of 170\,Hz, with the probe beam blocked (`Probe Off', measuring PL only), and unblocked (`Probe On', measuring $\Delta T$ and the dominant PL background). (c) Photoinduced absorption spectrum at 10\,K, measured with the double modulation apparatus (continuous line) and by single modulation after subtraction of the large PL background. We plot the absolute fractional change in transmittance ($|\Delta T/T|$), obtained by the $R$ output of the lockin amplifier. These data are acquired by an average of 25 successive scans with a time constant of 300\,ms.}
 \label{fig5}
 \end{figure}
 
 In order to demonstrate the power of the double modulation approach, we choose to measure the photoinduced absorption spectrum at 10\,K of a highly luminescent semiconductor polymer, poly(9,9-dioctylfluorene-co-benzothiadiazole), denoted F8BT~\cite{*[{See Table 1 in }] [{ for a summary of PL quantum efficiency values published for F8BT.}] Sardone:2007fk}. The structure of F8BT is shown in Fig.~\ref{fig5}(a). In the solid state at low temperature, F8BT has a PL quantum yield of $> 70$\%~\cite{Grey2006}. Neat F8BT films have been reported to feature a photoinduced absorption centred at $\sim 850$\,nm ($\sim 1.5$\,eV)~\cite{Westerling:2003ly}, assigned to absorption of triplet excitons from the lowest excited state to a higher-lying excited state, generated by intersystem crossing~\cite{Ford:2005ve,Lee:2007qf}. Fig.~\ref{fig5}(b) displays the signal measured with a single lockin amplifier upon modulating the pump laser at a frequency of 170\,Hz. The PL background dominates the $\Delta T$ signal, and is over a factor of 50 more intense than the features attributed to the photoinduced absorption of triplet excitons. The spectrum of $\Delta T/T$ obtained by subtracting the PL spectrum in part (b) before normalizing by the probe transmittance is shown as open circles in part (c). We also display the photoinduced absorption spectrum obtained by the double modulation scheme as a continuous line. The spectrum obtained by double modulation corresponds to that reported at room temperature~\cite{Westerling:2003ly,Ford:2005ve,Lee:2007qf}, where the PL background is significantly less problematic as the PL efficiency increases substantially at low temperature because of a decreased contribution of nonradiative decay channels such as internal conversion and diffusion-limited quenching at defect sites. On the other hand, the spectrum obtained by single modulation displays a substantial negative artifact at 752\,nm, resulting from substraction of a large PL background from the $\Delta T$ signal shown in part (b). At the red tail of the PL spectrum, the PL background may become comparable to the desired signal, and the subtraction procedure is very sensitive to the signal-to-noise quality of the data. It is in principle possible to achieve a signal-to-noise ratio that is adequate to subtract the PL background satisfactorily with sufficient signal averaging. However, the double modulation measurement as reported here is preferable given its simplicity and high intrinsic signal-to-noise capabilities because it is free from PL, laser scatter, and other spurious background sources. 
 
 We have estimated the ultimate detection sensitivity of our system. With an averaging time constant of 300\,ms in the lockin amplifier, we can measure a signal of $\sim 1$\,$\mu$V from the Si photoreceiver in low gain mode at frequency $f_{pu} + f_{pr} = 309$\,Hz, with the lockin sensitivity set to 10\,$\mu$V. Throughout much of the visible range, the probe signal at frequency $f_{pr} = 139$\,Hz is $\sim 1$\,V. We would thus have the capability of measuring a signal of $\Delta T / T \sim 1 \times 10^{-6}$, consistent with Fig.~\ref{fig4}(b). This is currently limited by the stability of our tungsten-halogen probe lamp, which is a grey body source with temperature stability limited by our current source. We estimate that with a more constant current supply, we would improve this sensitivity to $\Delta T / T \sim 1 \times 10^{-7}$.

\section{Conclusion}

We have demonstrated the use of a simple and inexpensive sum-frequency generation circuit to implement a double optical modulation apparatus that allows measurement of background-free photoinduced absorption spectra, and can also be implemented to measure photreflectance spectra. This circuit processes the TTL sync output signals from two independent chopper controllers to produce an output at the sum-frequency, which can be used as a reference signal for a lockin amplifier. Thus, the probe and pump arms of the experiment can be modulated at distinct frequencies $f_{pr}$ and $f_{pu}$, respectively, and the photoinduced absorption signal at $f_{pr} + f_{pu}$ can be measured. We have demonstrated the power of this apparatus by comparing the photoinduced absorption spectrum of a highly luminescent semiconductor polymer, F8BT, at 10\,K, where the luminescence background swamps substantially the pump-laser-induced change in probe transmittance. This double modulation instrument represents a simple and inexpensive means to measure high-signal-to-noise photoinduced absorption spectra of efficient emitters such as many organic semiconductors and inorganic semiconductor quantum structures, and is also well suited for photoreflectance spectroscopy.

\begin{acknowledgments}
This experimental apparatus was funded by a grant from the Natural Sciences and Engineering Research Council of Canada (NSERC), within the Research Tools and Instruments Scheme. CS acknowledges further funding from NSERC (Discovery Grant) and the Canada Research Chair in Organic Semiconductor Materials. HK acknowledges the Tunisian government for a graduate scholarship. GL acknowledges funding from the Royal Society for an International Travel Grant. GL and NS thank the European Community Seventh Framework Programme (FP7/2007-2013) for funding under Grant Agreement No. 212311 of the ONE-P project.
\end{acknowledgments}


%

\end{document}